\documentclass[]{aa}

\usepackage[dvips]{epsfig}
\usepackage{graphics}
\usepackage{amssymb}
\usepackage{times}

\newcommand{\field}[1]{\mathbb{#1}}

\newcommand{\R}{\field{R}}

\begin{document}

\raggedbottom

\parindent0em
\parskip1.5ex plus0.5ex minus 0.5ex

\renewcommand{\bottomfraction}{0.6}
\renewcommand{\baselinestretch}{1.1}

\def\3{\ss}
\def\re{\mathbb{R}}
\def\ze{\mathbb{Z}}
\def\ne{\mathbb{N}}
\def\pe{\mathbb{P}}
\def\ce{\mathbb{C}}
\def\ge{\mathbb{G}}
\def\Ee{\mathbb{E}}
\def\we{\mathbb{W}}
\def\Be{\mathbb{B}}
\def\Var{\mathbb{V}}
\def\se{\mathbb{S}}
\def\Covar{\mathbb{C}\mathbb{O}\mathbb{V}}

\def\se{\mathbb{S}}
\def\Covar{\mathbb{C}\mathbb{O}\mathbb{V}}
\def\U{\cal U}
\def\V{\cal V}
\def\vv{\bf V}
\def\uu{\bf U}
\def\kl{I \!\! [}
\def\kr{I \!\! ]}
\def \tieins{ t_{1, i+1} }
\def \ti{ t_{1, i} }
\def \tjeins{ t_{2, j+1} }
\def \tj{ t_{2, j}}

\newcommand {\indi}{ \mbox{ \bf I}_{[t_{1,i}, t_{1,i+1}) } }
\newcommand {\indj}{ \mbox{ \bf I}_{[t_{2,j}, t_{2,j+1}) } }

\def\trace{\thinspace {\rm tr}\thinspace}
\def\Q{{\cal Q}}
\def\diag{\thinspace {\rm diag}\thinspace}

\newcommand{\bea}{\begin{eqnarray*}}
\newcommand{\eea}{\end{eqnarray*}}
\newcommand{\be}{\begin{eqnarray}}
\newcommand{\ee}{\end{eqnarray}}
\newcommand{\wzbw}{\,\raisebox{-1.2ex}{$\Box$}}   
\newcommand{\indik}{\mbox{$1 \! \! \mbox{I}$}}
\newcommand{\sumi}{\sum\limits_{i=1}^{n}}
\newcommand{\sumk}{\sum\limits_{k=1}^{n}}
\newcommand{\suml}{\sum\limits_{l=1}^{n}}
\hyphenation{asym-pto-tisch}

\title{New statistical goodness of fit techniques in noisy inhomogeneous
inverse problems} 
\subtitle{With application to the recovering of the luminosity distribution
of the Milky Way}
\author{Nicolai Bissantz\inst{1} \and
Axel Munk\inst{2} }
\institute{Astronomisches Institut der Universit\"at Basel, Venusstr. 7, CH-4102 Binningen/Basel
\and Fakult\"at f\"ur Mathematik  und Informatik  der Universit\"at GH Paderborn,
Warburgerstr. 100, 33098 Paderborn}
\offprints{Nicolai Bissantz, Nicolai.Bissantz@unibas.ch}
\date{Received 16 June 2000 / Accepted 12 June 2001}
\titlerunning{Statistical Goodness of Fit Techniques}

\abstract{
The assumption that a parametric class of functions fits the
data structure sufficiently well is common in  
fitting curves and surfaces to regression data. One then derives
a parameter estimate resulting from a least squares fit, say, and
in a second step various kinds of $\chi^2$ goodness of fit measures,
to assess whether the deviation between data and estimated surface is due to 
random noise and not to systematic departures from the model. 
In this paper we show that commonly-used $\chi^2$-measures are invalid 
in  regression models, particularly  when inhomogeneous noise is present. 
Instead we present a bootstrap algorithm which is applicable in problems
described by noisy versions of Fredholm integral equations. of the first kind.
We apply the suggested method to the problem of recovering 
the luminosity density in the Milky Way from 
data of the  $DIRBE$ experiment on board the 
$COBE$ satellite. 
}
\maketitle

\section{Introduction}

Regression problems arise in almost any branch 
of physics, including 
astronomy  and astrophysics. In general, 
the problem of estimating a regression function (or surface)
occurs when a functional relationship between 
several quantities of interest has to be 
found from blurred observations $(y_i,t_i)$, $i=1, \cdots, n$.
Here $y =(y_1, \cdots, y_n)$ denotes a vector of 
measurements (response vector) and $t= (t_1, \cdots, t_n)$ 
a quantity which affects the  response vector in a 
systematic but blurred way  , which is to be investigated. 
This systematic component is usually denoted 
as the regression function $E[Y_i] = \omega(t_i)$. Note that 
$Y_i$ is a random variable, of which $y_i$ is a realisation.
If $t_i \in \re$, this includes 
signal detection problems  or image restoration if $t_i \in \re^2$. 
Many problems bear the additional difficulty that the quantity of interest 
is not directly accesible to the observations $y$  and 
the relationship has to  be expressed by a noisy 
version of a Fredholm integral eq. of the first 
kind, viz.
\be\label{modinv}
y_i = \omega(t_i) + \varepsilon_i = ({\bf K}\rho) (t_i) + \varepsilon_i,
\ee
where ${\bf K}$ is a given integral operator, $\rho$ the 
regression function to be reconstructed and $\varepsilon = 
(\varepsilon_1, \cdots, \varepsilon_n)$ a vector of independent 
 random quantities (error), due to imprecise  measurements  and other sources of 
noise. More precisely,  we assume that the expectation of $y_i$ is given by 
$({\bf K}\rho) (t_i)$ and inhomogeneous noise might be present, i.e. 
the variance  $\sigma_i^2$ 
 of the noise $\varepsilon_i$ (and possibly higher moments, too)  depends on 
the grid point $t_i$.
There is a vast amount of literature concerning  statistical theory for the
estimation of $\rho$, we mention only Wand \& Jones 
(\cite{wajo}) for direct regression and 
 Nychka \& Cox (\cite{nych}) or  van Rooij \& Rymgaart 
(\cite{rooi}) for the inverse (sometimes denoted as indirect)
 case, as in eq. (\ref{modinv}).
  (Inverse) regression  models capture various examples from astronomy  
and physics (cf. Bertero (\cite{berter}) or Lucy 
(\cite{lucy},\cite{lucy2}) for an overview). Such an example  is  
the reconstruction  
of the  three-dimensional
luminosity in  the Milky Way [MW], which will be discussed extensively 
in sect. \ref{mwappl}. In this example, $\rho$ will be a 
three-dimensional density
of the MW, ${\bf K}$ the operator that projects this density to the sky,
${\bf K}\rho (t_i)$ the resulting surface brightness at the sky position
$t_i=(l,b)_i$ and $y_i$ the observed surface brightness at $(l,b)_i$.

Reconstruction  procedures  (estimation) of $\rho$ in general depend 
on various a priori assumptions about $\rho$, such as smoothness properties
or geometrical constraints, e.g. monotonicity. 
The most common assumptions are that $\rho$ has  a particular structure and
shape, depending on some unknown parameter $\vartheta$. Such an assumption is  
denoted as a parametric model. Typically, these strucutural
assumptions arise from  physical reasoning and approximation procedures.
Often, however, it is not completely clear whether 
 these assumptions are satisfied and therefore it is an important task
to investigate empirically (by means of the data at  hand) whether the resulting model
is valid. Therefore, in this paper we  discuss recent methodology
for the investigation of  the adequacy of such a parametric  model 
$U= \{ \rho_{\vartheta} \}_{\vartheta \in \Theta}$, 
$\Theta \subset \re^d$. This will be done 
for regular  regression  problems  as well as for the inverse case, as in (\ref{modinv}).  
 
The paper is organized as follows. 
In the next sect. we briefly review common practices to judge the 
goodness of fit of a model  $U$. It is  shown that 
classical goodness of fit approaches, such as least square statistics 
are insufficient from  many  methodological points of view, particulary when 
inhomogeneous noise is present, i.e. the variation $\sigma_i^2$ of the error 
$\varepsilon_i$  is expected to vary with the grid point (covariate) $t_i$. 
We show in section 2 that statistically valid conclusions about the goodness of fit 
from the residuals  $\sum r_i^2=\sum \left(y_i-(\bf K \rho_{\hat\theta}) (t_i)\right)^2$ 
(or variants of it) are impossible
in general, particularly when inhomogeneous noise is present, as is the case in 
our data example. This is mainly due to the fact that in the inhomogeneous
case the distribution of $\sum r_i^2$ depends on the whole vector 
$\left(\sigma_1^2,\ldots,\sigma_n^2\right)$ which is in general unknown.
Therefore, we suggest in sect. 3 a measure of fit which is based on ``smoothed residuals''
and which allows for the calculation of the corresponding probability distribution.
In sect. \ref{bootstrap}, a bootstrap resampling algorithm 
is suggested which allows the algorithmic reconstruction of the distribution 
of the suggested goodness of fit quantity.
The use of bootstrap  techniques is well documented in astronomy (cf. Barrow et al. 
(\cite{barrow}), Simpson and Mayer (\cite{simpson}), van den Bergh  and Morbey 
(\cite{vandenbergh}) for various applications). The work similar in spirit to ours 
is Bi \& B\"orner's  (\cite{biboerner}) residual type bootstrap, used as a method for 
nonparametric estimation in inverse problems. As a byproduct we show, however, that 
this residual bootstrap is insufficient in the case of inhomogeneous noise in the data
and a so-called 'wild' bootstrap has to be used instead.

Finally we will apply our new method in sect. 5 to the fit of the {\it COBE/DIRBE} L-band 
data. We use a functional form for a parametric model of the MW as presented
by Binney et al. (\cite{binney96}, hereafter BGS) and find similar structural 
parameters 
of the Milky disk and bulge, except for the scale height of the disk which we find to
be about $25\%$ smaller. 

\section{Common $\chi^2$ methods of judging the quality of fit}
One of the most popular techniques for finding a
proper fit of a given model $U$  to a given set of data
$y_1, \cdots, y_n$ is to
minimize  a (penalized) weighted sum of squares
\bea
Q_{\tilde w}^n (\vartheta):= \sumi {\tilde w}_i  (y_i - \omega_{\vartheta}(t_i))^2
\eea
where the $\tilde w_i$ denotes some weighting scheme and
the model is $\omega_{\vartheta}(t_i)=({\bf K}\rho_{\vartheta}) (t_i)$. 
This leads to a weighted least squares
estimator (WLSE) of the optimal model parameter, $\hat \vartheta_{\tilde w}$.
However, it is well known that a proper choice of the weights
$\tilde{w}_1, \ldots, \tilde{w}_n$ depends on the (possibly position-dependent) 
random noise in the data. For example, under an uncorrelated  normal error assumption,
if the variance $ \sigma_i^2$ of the error
$\varepsilon_i$ is assumed to  be known, a suitable choice of weights is $w_i=\sigma_i^{-2}$
in order to take into account the
local variability of the observations at the grid point $t_i$.
Particularly, in  this case, the ordinary, unweighted least squares estimator
is known to be insufficient (Gallant \cite{gall}), because the log likelihood
of the model is proportional to $Q^n_{\tilde w}(\vartheta)$.
Only if the variance pattern is homogeneous (i.e. $\sigma_i^2 = \sigma^2$)
are unweigthed least squares estimators optimal. 
The weighted least squares  approach is, however, limited if the 
the local variances $\sigma_i^2$ in the data points are unknown. 
The $\sigma_i^2$ then have to be estimated from the data. This is often
neglected.  
It is also common practice to consider  $Q^n_{w}$ in order to judge the 
quality of fit achieved by the regression function (where the weights $w_i$ may 
sometimes be different to those used in computing the WLSE).
Here, a 'large'  value of $Q^n_{w}$ is used as an indicator for  a 'significant' deviation  
between the observations and the  model to be fitted. We will investigate this
in more detail in what follows, and emphasize the case of 
nonhomogeneous variances.

In general, the most important properties required
of any goodness of fit (GoF) quantity $\chi^2$ such as $\chi_w^2$ are that 
\begin{enumerate}
\item[1)] we are  able to detect with high probability deviations from the model
 we have in mind (often denoted by statisticians  as good 'power'),
\item[2)] we can quantify the probability that $\chi^2$ exceeds some "critical value" in order to obtain
a precise probabilistic analysis (computing significance levels, confidence intervals, etc.).
\end {enumerate}
As a rough rule of thumb often 
\be
\frac {1}{n-d} \; \chi^2_w \approx 1
\label{rough}
\ee
is taken as a measure of evidence for the model {\it U} and hence for the 
fit of $\omega_{\hat{\vartheta}_{\tilde{w}}}$. Here  {\it d } 
denotes the dimension  (number of parameters) of {\it U}  and {\it n}  the number of 
data points. Examples of the use of this kind of statistics can
be found in Alcock et al. (\cite{alcock}) and 
in Dwek et al. (\cite{dwek}) in the context  of discriminating between several models.
A related well-known quantity is the sum of squares of
'expected minus observed divided by expected'
for testing 
distributional assumptions, such as normality  of the data (cf. Cox \& Hinkley \cite{cohi}).
Bi \& B\"orner (\cite{biboerner}) considered a similar quantity 
in a deconvolution setup  which is, using  the notation of (\ref{modinv})
\be\label{genausoblind}
\sumi \frac{\hat r_i^2}{\omega_{\hat \vartheta}(t_i)}.
\ee
This  obviously downweights the influence of residuals if the corresponding predicted value
$\omega_{\hat \vartheta} (t_i)$ is large. Another option is to consider the absolute deviation of
the predicted and observed values, which leads to a more robust version of $\chi^2$, or
even more general distance measures can be used (Cook \& Weisberg \cite{cowei},
Hocking \cite{hock}, Lucy \cite{lucy}, \cite{lucy2}). 
In the following we will argue that an approach like $\chi^2_w$ is not valid in regression 
models such as (\ref{modinv}), particularly when the noise is inhomogeneous or the 
residuals are not gaussian. To this end we briefly discuss the (asymptotic) 
distribution of the abovementioned quantites.

In order to get a first insight into the probabilistic behaviour of statistics such as
$\chi^2$ , used as a quantitative measure of fit, it is helpful to consider
the distribution in the simplest case when $\omega\equiv 0$. A simple calculation then shows 
that (assuming a normal distribution
of the data) $\chi^2 = \sumi y_i^2$ is distributed as a sum of normally distributed variables having the
expectation $E[\chi^2] = \sumi \sigma_i^2$, and  variance $V[\chi^2] = 2\sumi \sigma_i^4$.
Hence, already in this simple case it can be seen that the determination of the law of
$\chi^2$ is practically impossible if the variances $\sigma_i^2$ are 
not known. Then it is difficult to quantify what a "too large value of
$\chi^2$" means, because this will depend on the unknown quantities
$\sigma_1^2, \ldots, \sigma_n^2$, and a rule as in (\ref{rough}) can lead in
principle to any result in favour or against the model $\omega \equiv 0$.
We mention that standardisation by the predicted values as in (\ref{genausoblind})
does not avoid this problem. This is in 
contrast to  goodness of fit problems
for the assessment of distribution assumptions, i.e. 
when one investigates by a $\chi^2$ measure whether
a population is normal, say (Cox \& Hinkley \cite{cohi}). 
Note, that the case of homoscedastic regression models (i.e. the distribution of the
noise is identical for all data points) is  somewhat simpler, because here the expectation
$E[\chi^2]=n\sigma^2$ and the square root of the variance $V[\chi^2]= 2n\sigma^4$
is proportional, i.e. the signal to noise ratio
\bea
\frac{E[\chi^2]}{\sqrt{V[\chi^2]}} = \sqrt{n/2}
\eea
only depends on the number of data points $n$. Here, a model-free estimator of $\sigma^2$
can be used as a reference scale (Hart \cite{hart}). 

Many attempts were 
made in order to find simple approximations of the  distribution for
$\chi^2_w$. Among them 
a quite attractive option is use of a bootstrap method, an algorithmic
approximation of the true law (see Efron \& Tibshirani \cite{ef2}
for an overview  and many applications).
Bootstrapping random quadratic forms (such as $\chi^2$)
is, however, a rather delicate matter, because standard 
bootstrap algorithms such as Efron's  (\cite{ef1}) $n$ out  of $n$ bootstrap are inconsistent 
 (Babu (\cite{babu}), Shao \& Tu (\cite{shaotu})), i.e.  the distribution 
is not approximated correctly  with increasing number of observations. 

The use of a particular bootstrap algorithm
 was indeed suggested by Bi \& B\"orner (\cite{biboerner})  in the context
of assessing the goodness of fit 
in deconvolution  models. We mention that their bootstrap
algorithm, however,  is  asymptotically not correct
in inhomogeneous models.
Interestingly, the suggested algorithm is similar  in
spirit to the  so called residual bootstrap (i.e. drawing random samples with
replacement from the residuals $r_i$) which is well documented
in the statistical literature (cf. Davison \& Hinkley (\cite{davhi}) p.281) for
the estimation of the regression parameters).\\

Despite the abovementioned difficulties, 
the main problem encountered with the naive use of 
$\chi^2$ in regression models as a measure of GoF is 
 that asymptotically (here and in the following, asymptotically means
the sample size tends to infinity) the law of $\chi^2$ does in
general not converge asymptotically to any reasonable quantity, in contrast 
to goodness of fit testing for distributional assumptions. Even after
rescaling by $1/ \sqrt n$ in order to force  the variance
\bea
V[1/ \sqrt n \; \chi^2] = 2/n \sumi \sigma_i^4 \;
\stackrel {n \rightarrow \infty}{\longrightarrow} \;
 2 \int \sigma^4 (t) H(dt)
\eea
to  converge (here it is assumed 
that  the scheme of grid points can be 
described asymptotically by a  distribution $H$ (Dette \& Munk \cite{demu1})) gives
\bea
E[1/ \sqrt n \; \chi^2] = \sqrt n ( 1/n \sumi \sigma_i^2)
\, = O(\sqrt{n}) \, \rightarrow
\infty
\eea
which shows that $1/ \sqrt n \; \chi^2$ does not converge to any reasonable quantity.
Note that if we use $\chi^2/n$ the variance tends to zero. 
Also observe that subtracting $E[1/ \sqrt n \; \chi^2]$ from $\chi^2$ 
will not provide a way out of the dilemma because this value depends
on  the (unknown) variances $\sigma_i^2$. 

In summary, we see that without explicit 
knowledge of the variances $\sigma^2_i$, 
the use of $\chi^2$ as a {\it quantitative}
measure of validity  of a model is not appropriate.

Due to the above-described difficulties, statisticians throughout the last  two  decades 
have extensively studied the
problem of checking the goodness of fit in regression models. 
It is beyond the scope of this paper to review
this work; many references can be found in the
recent monograph by Hart (\cite{hart}). Among the variety of 
procedures suggested so far, we mention methods
which are based on model selection criteria,
such as Akaike's (\cite{akai}) information criterion
(Eubank \& Hart (\cite{euha}), Aerts et al. (\cite{aerts}))
and methods which compare nonparametric estimators with a parametric
estimator. To this end Azzalini \& Bowman (\cite{azzbow}), 
H\"ardle \& Mammen (\cite{haermam})
and M\"uller (\cite{muel}) used a kernel
 estimator, Cox et al. (\cite{cowa}) smoothing splines and
Mohdeb \& Mokkadem (\cite{mohdmok}) a Fourier series estimator.
However, the applicability of many of these methods is 
often limited. For example, H\"ardle \& Mammen's
test is confronted with bias problems, whereas
other procedures are only applicable  for  homogeneous
errors or when the error distribution is completely known
(Eubank \& Hart \cite{euha}, Aerts et al. \cite{aerts}).
Another serious difficulty arises with the nonparametric estimation of 
the signal as the dimension {\it k} of the grid points increases. This
is sometimes denoted as the curse of dimensionality (Wand \& Jones  \cite{wajo}, 
Bowmann \& Azzalini \cite{boazz}).
A rough rule of thumb is that the number of observations required
in dimension $k$ is $n^k$ in order to obtain the same
precision of   the estimate of $\omega$.  Hence, the precision
induced by  100 observations on the real line is approximately the same as
10,000 drawn from the plane.  Furthermore,  
measurements often cannot be taken equidistantly over a grid,  which
leads to sparse data structures causing further difficulties
with increasing dimension.
One should also note that another difficulty consists of
transferring these methods to the case of inverse
problems, a situation which up to now has never been treated.

\section{A new method}
Munk \& Ruymgaart (1999)
have developed a general regression methodology which remains valid in 
the heteroscedastic case (i.e. the distribution of the noise depends on 
the data point) with arbitrary dimensions of the grid points. The
underlying idea dates back to H.Cram\'er  and
can be summarized as "smoothing the residuals"
in order to obtain asymptotical stabilization of
the test criterion. In our context this reads
as follows. Let {\bf T} denote an injective smoothing linear 
integral operator with associated integral kernel $T(\cdot, \cdot)$, i.e.
\be
{\bf T}[f](u) = \int T(u, t) f(t) dt.
\label{inveq}
\ee
Note that since $\bf T$ is an integral operator, ${\bf T}[f]$ is again
a function. Now consider the transformed
distance between the parametric model 
$V=\{\omega_{\vartheta}\}_{\vartheta\in\Theta}$ 
and the distribution $\omega$, which underlies the observations 
$y_i=\omega(t_i)+\varepsilon_i$ (cf. sect. 1), 
\be
D^2 (g) = D^2({\bf T}[\omega]) = \min_{\vartheta \in \Theta}
\| {\bf T} [\omega-\omega_ {\vartheta}] \|^2
\label {trans}
\ee
where $g = {\bf T}[\omega]$ denotes the smoothed version 
of $\omega$ and  the norm $ \|. \| $ refers to some
$L^2$-norm to be specified later on. 
The smoothed distance $D^2$ serves now as a new measure
of goodness of fit and has to be estimated from data.
This will be done by numerical minimization of
the empirical counterpart of the r.h.s. of (\ref{trans}), 
$\| \hat g- g_{\vartheta} \| ^2 = \chi^2_D
(\vartheta)$ where $\hat g (u) = 1/n \sumi y_i \; T(u, t_i)$
denotes an estimator of $g(u)$. In addition this provides
us with a smoothed estimator  $\hat \vartheta_{{\bf T}}$
of the value $\vartheta_{{\bf T}}^{\ast}$ for which the minimum 
in (5) is achieved. 
In Munk \& Ruymgaart (1999) 
the  kernel
\bea
T(u,t) = \min (u,t), \qquad
\eea
was suggested (see Appendix A1), which will also be used in the following,
and which amounts to a cumulative smoothing. 
Not that for
{\it k}-dimensional {\it u} and {\it t} the minimum has to be
understood componentwise as
\bea
\min(u,t) = \prod_i^k \min (u_i, t_i) \qquad.
\eea
We mention that other choices of $T$ are possible (cf. Sect. \ref{remarks}). 

The reasoning behind this approach is that
direct estimation of $\omega$ is a rather difficult
task, whereas estimation of the smoothed
transformation $ g= {\bf T}(\omega)$ is much simpler. 
Furthermore, the distribution of the minimizer of $\chi^2_D(\vartheta)$
becomes tractable. 
If we denote the minimum of $\| \hat g -
g_{\vartheta} \| ^2$ as $\hat D^2$ one can
show under very mild regularity conditions
that  (Munk \& Ruymgaart, 1999) the distribution of
$n \hat D^2$ converges to 
\be
\sum_{i=1}^{\infty} \lambda_i \chi_i^2,
\label {lim}
\ee
where the $\lambda_i$ denotes a decreasing
sequence of positive numbers which depend on the best model parameter
$\vartheta^{\ast}_{\bf T}$, which is the minimizing $\vartheta$ in (\ref{trans}), 
the model space $U$ and the unknown distribution
of the error $\varepsilon$, including the variance function
$\sigma^2(t_i)$. This makes a
direct application of this limit law
difficult and hence a peculiar bootstrap algorithm
is suggested in the following which can be
shown to be asymptotically consistent, i.e.
the asymptotic limit law of this algorithm is the same as in
(\ref {lim}). The following idea of the so called
"wild bootstrap" dates back to Wu (\cite{wu}) and
was applied by Stute et al. (\cite{stut2}) to a 
testing problem similar to the one above.

\section{The wild Bootstrap Algorithm}
\label{bootstrap}
The true distribution (6) of $\hat D^2$ depends on the 
unknown $\lambda_i$. It is therefore not possible to use
this distribution for practical purposes. 
However, it is possible to approximate the
distribution numerically using the following bootstrap 
algorithm:

{\bf Step 1:}  ({\it Generate residuals}). Compute residuals
\bea
\hat \varepsilon _i
:= y_i-\omega_{\hat \vartheta_{\bf T}}
(t_i), \qquad i=1, \cdots, n
\eea
where ${\hat \vartheta_{\bf T}}$
denotes a solution of the
minimization of
\bea
 \hat D^2 := \chi^2(\hat \vartheta_{\bf T})
:= \min_{\vartheta \in \Theta}
\| \hat g -  {\bf T}\omega_{\vartheta} \|^2.
\eea

\begin{figure}
\resizebox{\hsize}{!}{\includegraphics{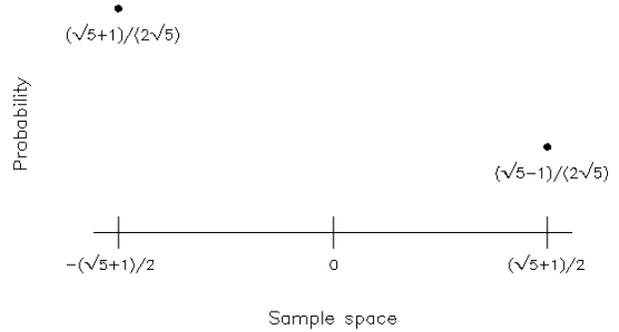}}
  \caption{Binary probability distribution required in step 2
of the wild bootstrap algorithm.}
\label{figprob}
\end{figure}

{\bf Step 2:} ({\it The "wild" part}).
Generate  new random variables $c_i^{\ast}, \; i=1, \ldots, n$, 
which do {\it not} depend on the data, where each $c_i^{\ast}$
is distributed to a distribution
which assigns probability
$(\sqrt{5}+1)/2 \sqrt{5} $ to the
value $(- \sqrt{5} -1)/2$ and
$(\sqrt{5} - 1)/2 \sqrt{5}$ to the
value $(\sqrt{5} + 1)/2$. See fig. \ref{figprob} for
a visualization of this probability distribution.

{\bf Step 3:} ({\it Bootstrapping residuals}). Compute
$\varepsilon_i^{\ast} := \hat
\varepsilon_i c_i^{\ast}$ and $y_i^{\ast} = 
\omega_{\hat \vartheta_{\bf T}}
 + \varepsilon_i^{\ast}$.
This gives a new data vector 
$(y_i^{\ast}, t_i)_{i=1, \ldots, n}$.

{\bf Step 4:} ({\it Compute the target}).
Compute $\hat D^{2, \ast}$
with $(y_i^{\ast}, t_i)_{i=1, \ldots, n}$

{\bf Step 5:}  ({\it Bootstrap replication}). Repeat step 1-4 {\it B}
times ($B=1000$, say) which gives
values $\hat D_1^{2,\ast}, \ldots,
\hat D_{B}^{2, \ast}$.\\

Now we construct the empirical cumulative distribution function [ECDF], which
can be taken as an approximation for the right side in (\ref{lim}), because 
Munk \& Ruymgaart (1999)
have shown that the ECDF, based on $\hat D_i^{2,\ast}$, 
asymptotically approximates the distribution 
of $\hat D^2$. The ECDF can be obtained by ordering the
values of $\hat D_1^{2,\ast}, \ldots,
\hat D_{B}^{2, \ast}$ increasingly and plotting them
against the value $(i)/B$, where
({\it i}) denotes the position of
$\hat D_i^{2,\ast}$ in the ordered sample
$\hat D_{(1)}^{2,\ast}, \ldots,
\hat D_{(B)}^{2, \ast}$. The so-called estimated
evidence  of the model $U$ can now be obtained by determining
the position of the original statistic
$\hat D^2$ in the ordered sample
$\hat D_{(1)}^{2,\ast}, \ldots,
\hat D_{(B)}^{2, \ast}$. This
is some number $k^{\ast} \in \{ 0,
\ldots ,B+1 \}$. From this number one computes 
\bea
\alpha^{\ast} = 1- k^{\ast}/B.
\eea
Statisticians denote 
$\alpha^*$  as the $p$-value of the test statistic $\hat D^2$.
The interpretation of this value is as follows. A small $\alpha^*$
indicates that the  observed data are very unlikely to have been 
generated by  model $U$, because the probability that 
the observed (or a larger value) $\hat D^2$ occurs is very small, 
namely $\alpha^*$  (recall that the bootstrap algorithm 
reproduces the true distribution of $\hat D^2$ in (\ref{lim})).  
On the other hand, if $\alpha^*$ is large (and hence $\hat D^2$ small)
 there should be 
rare evidence against a proper use of model $U$, because 
$\hat D^2$ provides a good fit of the data to the model compared 
to all other possible outcomes which could have occured.  

A formal test at significance level
$\alpha$ can be performed
when deciding against {\it U} if
\be
\alpha^{\ast} <\alpha.
\label{eqsig}
\ee
In other words a small $\alpha^{\ast}$
indicates that the deviation from the model
$U$ is not simply due to noise, but rather a
systematic devation from the model  $U$
has to be taken into account.

We would like to close this section by making some remarks
about the applicability of bootstrap algorithms  in the 
context of goodness of fit, and giving some arguments why
our bootstrap algorithm is valid in the heteroscedastic case.
Stute et al. (\cite{stut2}) have shown that the wild  bootstrap is
valid in heteroskedastic models with
{\it random} grid points $t$. 
This result can be extended to
deterministic grid points, as is the case in our example, 
provided the scheme is not  'too'
wiggly  (a precise formulation can be found in
Munk (\cite{mu1})), which holds true for the subsequent example.
We mention that an explanation for the wild bootstrap validity
is its automatic adaptivity to inhomogeneous variances, because it 
can be shown that the variance in the artifical datapoints $y^{\ast}_i$
induced by ``wild'' resampling (step 2 in our algorithm) yields
\bea
V^{\ast}[\varepsilon^{\ast}_i] = \hat\varepsilon_i^2
\eea
which estimates approximately $\sigma^2(t_i)$. In contrast, the $n$ out
of $n$ bootstrap (cf. Stute et al. 1998) and the residual bootstrap 
here fail to hold because the bootstrap variance is in the latter
case $\frac{1}{n} \sum_{i=1}^{n} \hat\varepsilon_i^2$, which approximates
the average overall variance $\int\!\!\int \sigma^2(t_1,t_2)dt_1 dt_2$ in our example. 
This argument transfers essentially  to any random quadratic form 
(such as $\hat D^2$ or Bi \& B\"orner's (\cite{biboerner}) $\chi^2$-statistic). 
The residual bootstrap is consistent 
only if the error is homoscedastic, which, however, 
in the subsequent example is not the case.
The case when the model space $U$ is
of dimension $\infty$, as
considered by Bi \& B\"orner (\cite{biboerner}), is in principle similar;
here  it is also well known 
that   the residual bootstrap is insufficient in heteroscedastic models
(H\"ardle \& Marron \cite{haermar}).

\section{Application: recovering the luminosity distribution in the Milky Way}
\label{mwappl}

The {\it DIRBE} experiment on board the {\it COBE} satellite, 
launched in {\it 1989},
made measurements of the surface brightness in several infrared wavebands
(Weiland \cite{weiland94}). A difficulty with the  {\it COBE/DIRBE} 
data is that it has to be corrected against certain effects.
The most important correction
is the removal of dust absorption. This has been done by Spergel et al. (\cite{spergel95}). 
We use their corrected {\it COBE/DIRBE} L-band data in our fits. 
The resolution of the data are $n\!\times\! m\! =\! 120\!\times\! 40$ points
in $l,b$ respectively, covering a range $-89.25^o\! \leq\! l\! \leq \! 
89.25^o$ and $-29.25^o \! \leq\! b \!\leq\! 29.25^o$. 
The points in this two-dimensional grid are equally spaced. 

The {\it COBE/DIRBE} data have been used to deproject the three-dimensional density of the MW
in a number of projects. A main difficulty in 
recovering the three-dimensional luminosity distribution from the two-dimensional
surface brightness distribution of the MW is that 
it is not a unique operation, in general.
One way to avoid this problem is to fit a parametric model to the MW in 
order to reduce the set of possible models. 
Several  parametric models have been suggested,
see for example Kent et 
al. (\cite{kent}), Dwek et al. (\cite{dwek}) or Freudenreich (\cite{freudenreich}). Another  approach is
to use the non-parametric Richardson-Lucy 
algorithm for the deprojection of the data 
(Binney \& Gerhard \cite{binney95}, Binney et al. \cite{binney96}, Bissantz et al. \cite{bissantz97}),
in order to reconstruct the luminosity distribution of the MW.

In parametric models of the MW density, about  ten ``structural
parameters'' - including normalisations, scale lengths and geometrical shape
parameters of the bulge/bar - are used (see, for example, Kent et al. \cite{kent}, Dwek et al. \cite{dwek}
\& Freudenreich \cite{freudenreich}). 
In what follows, we assume that these parameters are selected
such that the projection of a model onto the sky is an injective operation.

\subsection{The basic eqs. of the astrophysical problem}

\begin{figure}
\resizebox{\hsize}{!}{\includegraphics{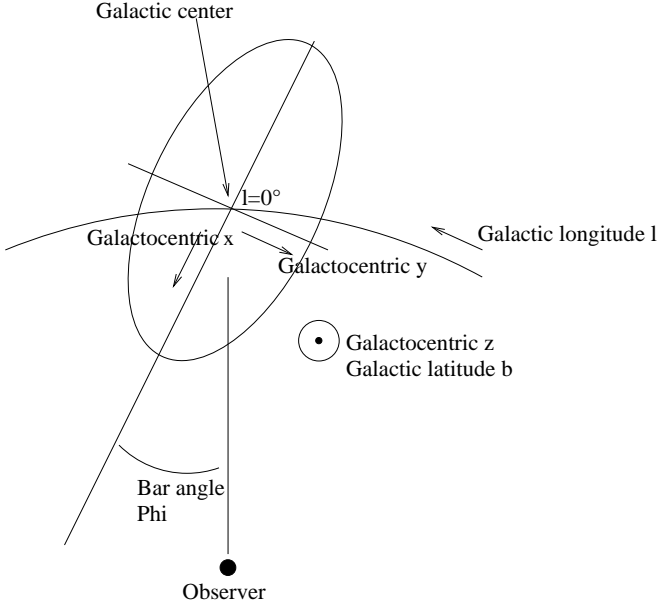}}
  \caption{A sketch of the two coordinate systems that we use in this
paper. Luminosity densities of the MW are defined in the galactocentric
coordinate system $x,y,z$. Galactic longitude $l$ and latitude $b$
define a position on the sky. Together with the distance from the observer
$r$ they constitute the observer centered coordinate system.}
\label{figcoo}
\end{figure}

We will first derive a general
mathematical model of the problem of recovering the MW luminosity
distribution from the L-band data. 
The projection of a three-dimensional light distribution 
to a surface brightness (on the sky) is defined as follows.
Let $\aleph$ be the set of possible luminosity densities of the MW, i.e.
of maps
\bea
\rho: \R^3  \rightarrow  \R_{\geq 0}, \qquad
 \vec{x}  \mapsto  \rho\left(\vec{x}\right)
\eea
and $\Omega$ be the surface brightness distributions
\bea
\omega: \left[0,2\pi\right]\times\left[-\frac{\pi}{2},\frac{\pi}{2}\right]  
\rightarrow  \R_{\geq 0}, 
       (l,b)  \mapsto  \omega (l,b), 
\eea
where $\omega(l,b)$ is the surface brightness at sky position $(l,b)$,
and $\omega\in\Omega$.
The transformation between 
a luminosity density 
$\rho$ to its corresponding 
  surface brightness distribution
is described by 
a linear integral operator ${\cal P}$.  We will call this operator
${\cal P}$ the projection operator, since it ``projects'' a luminosity
density on the sky, i.e. onto a surface brightness distribution.
\bea
{\cal P}: \aleph  \rightarrow  \Omega, \qquad
\rho & \mapsto & \cal{P} \left(\rho \right)
\eea
${\cal P}$ is defined  by the integral $\int \rho(\vec{x}) dr$
of the density $\rho(\vec{x})$ 
along the line-of-sight from the observer to infinity in
direction $(l,b)$. Let $r$ denote the distance from the 
observer. Note that the integrand is $\rho$ and not $\rho \times r^2$
because the physical extend of the observed cone $\delta\Omega$
increases as $r^2$ whereas 
the intensity of a source decreases as $r^{-2}$ and the $r$-powers
therefore cancel out.
Let $\vec{s}(r,l,b)$ be the path from the observer to infinity
along the line-of-sight to $(l,b)$, parametrized by
the distance from the observer $r$. Then
\bea
\omega(l,b) = {\cal P}\left(\rho\right) (l,b)  =
\int_{0}^{\infty} \rho(\vec{s}(r,l,b)) dr.
\eea
So far we have used a spherical coordinate system centered at the observer. 
Coordinate axes are the sky longitude $l$ and latitude $b$ and the 
distance from the observer $r$.
We now introduce a second coordinate system. 
Let $(x,y,z)$ be the coordinate axes of a  cartesian, galactocentric,
coordinate system, s.t. $x$ and $y$ lie in the main plane of the MW. We define
$x$ to be along the major axis of the bulge/bar (cf sect. \ref{model} for 
further explanation of the components of our parametric MW model, including 
the bulge/bar), $y$ along the minor axis, and $z$  
perpendicular to the main plane of the MW. 
Let us further call ``bar angle $\Phi$'' the angle between the major axis 
of the bar/bulge and the line-of-sight direction from the observer to the 
galactic centre. The position of the observer in this coordinate system is
denoted as $(x_{\odot},y_{\odot},z_{\odot})$.
Fig. \ref{figcoo} depicts the two coordinate systems. 

In our setting, a parametric model of the 
MW is a class $U$ of distributions $\rho_{\vartheta}$ 
such that
\bea
U=\left\{\rho_{\vartheta}(\cdot)\right\}_{\vartheta\in\Theta}
\eea
where $\Theta$ denotes a 
set of parameters $\Theta\subseteq\R^d$. This definition is 
in accordance with our terminology at the beginning of this paper. 
We cannot observe $\omega$ directly, due to measurement errors, 
dust removal from the raw data and other sources of noise. 
Hence, our observations $y_{ij} \equiv \omega(l_i,b_j)+\varepsilon_{ij}$ 
are blurred by some random error $\varepsilon_{ij}$, the distribution 
of which may vary between different sky positions $(l_i, b_j)$. 
Particularly, we  will see in the following that it is necessary to
allow for a position-dependent noise 
$Var[\varepsilon_{i,j}]=\sigma^2_{i,j}$. Therefore our astrophysical problem is
to reconstruct $\rho$ from the noisy integral eq. 
\bea
y_{ij} = {\cal P}({\rho})(l_i,b_j) + \varepsilon_{ij}.
\eea
Note that  this is not a noisy Fredholm eq. of the first kind as
in (\ref{inveq}); 
however the suggested method in the last section
transfers directly to the present setting.
Note, that $\cal P$ is   a linear injective operator  as long as
$\rho > 0$ due to our selection of the parametric model.  

Let  
\bea
\omega_{\vartheta}(l,b) ={\cal P}\left({\rho}_{\vartheta}\right)(l,b); \qquad \vartheta \in \Theta,
\eea
and consider the transformed model 
\bea
V_{\bf T}  = {{\bf T}V}=\left\{{\bf T}\omega_{\vartheta}(l,b)  \right\}_{\vartheta \in\Theta},
\eea
where
\bea
V={\cal P}U=\left\{\omega_{\vartheta}(l,b)\right\}_{\vartheta\in\Theta}
\eea
and
\bea
\textrm{{\bf T}} [\omega](l',b') = \int\int \omega(l,b) T((l',b'), (l,b)) dldb.
\eea
Specific models $U$  will be discussed in the next sect.
Following the approach in 
sect. \ref{bootstrap} we  specify the smoothing 
integral operator {\bf T} by defining the smoothing kernel as
\bea
T((l',b'),(l,b))=\min\{l,l'\}\cdot \min\{b,b'\}; \,\, (l,b),(l',b') \in \R^2.
\eea
This amounts to a kind of cumulative smoothing, which downweights small-scale
features in the data, and emphasizes trends on large scales. 

Now, as a first step, we estimate $g(l',b')=\textrm{{\bf T}}[\omega](l',b')$ 
by 
\bea
\hat{g}(l',b')=\frac{1}{n\cdot m} \sum_{i=1}^n \sum_{j=1}^m y_{ij} T((l',b'),(l_i,b_j))
\eea
and determine numerically the 'transformed' LSE
\be\label{best}
\hat{\vartheta}_{\bf T} 
= \textrm{argmin}_{\vartheta\in\Theta} ||\hat{g}-{\bf T}[\omega_{\vartheta}]||^2
\ee
where $||\cdot||^2$ denotes the  usual $L^2$-norm.
Finally, the minimising value
\bea
\hat{D}^2= ||\hat{g}-{\bf T}[{\cal P}\left(\rho_{\hat{\vartheta}_{\bf T}}\right)]||^2
\eea
is computed. Now the bootstrap algorithm in sect. \ref{bootstrap}
can be applied.
Finally, we mention that for  the minimization in (\ref{best})
we have used the Marquardt-Levenberg algorithm (Press et al. \cite{press}). 

\subsection{A parametric model of the Milky Way}
\label{model}
We will now investigate 
whether  the functional form of the luminosity distribution 
of the MW as suggested by BGS 
provides a satisfactory fit to the {\it COBE/DIRBE} L-band data. 
This functional form is a superposition of a double-exponential disk with a truncated
power-law bulge
\begin{eqnarray}
\rho_{\vartheta}(x,y,z) & = & \underbrace{d\cdot\left( \frac{e^{-|z|/z_0}}{z_0}
 + \alpha \cdot \frac{e^{-|z|/z_1}}{z_1}\right)\cdot r_d \cdot
e^{-r/r_d}}_{\textrm{disk}} \nonumber\\
 & & +\underbrace{b\cdot\frac{e^{-a^2/a_m^2}}{a^3_m\cdot
\eta\cdot\zeta\cdot\left(1+a/a_c \right)^q}}_{\textrm{bulge/bar with cusp}} \nonumber\\
a^2 & \equiv & x^2+\left(\frac{y}{\eta}\right)^2 +
\left(\frac{z}{\zeta}\right)^2 \nonumber\\
r^2 & \equiv  & x^2+y^2, 
\label{disk}
\end{eqnarray}
where the parameter $\vartheta$ 
can be devided into ``structural'' parameters 
$(z_0,z_1,r_d,b,a_m,\eta,\zeta,a_c,d,q)$ that specify
the functional form of the model and ``geometrical'' parameters that
define the position of the sun in the coordinate system.
The ``geometrical'' parameters are fixed in advance and are the position
of the sun above the main plane of the MW,
$z_{\odot}=14pc$, the distance of the sun from the galactic centre projected
on the main plane, $r_{\odot}=8kpc$ and the bar angle, $\phi=20^o$
(BGS). It is not   feasible
to estimate the cusp parameters $a_c=100pc$ and $q=1.8$ 
from our data, because the available resolution is not sufficient. We use
the same values as BGS.

\begin{figure*}
\resizebox{\hsize}{!}{\includegraphics{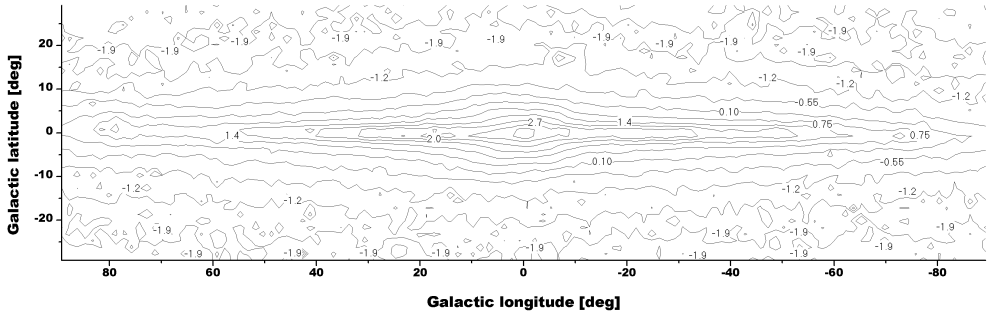}}
  \caption{{\it COBE/DIRBE} L-band data. Contours levels are given in 
magnitudes. Note that contour levels are only defined up to a common
offset.}
  \label{figcobe}
\end{figure*}

\begin{figure*}
\resizebox{\hsize}{!}{\includegraphics{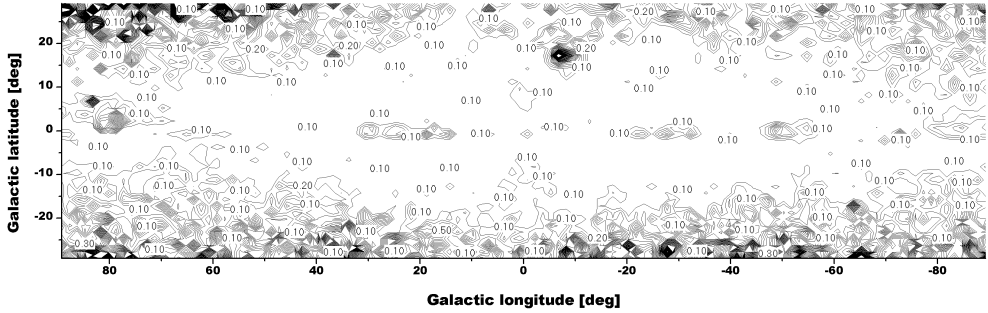}}
  \caption{ Square difference between {\it COBE/DIRBE} L-band data logarithmic
surface brightness (magnitudes) and the parametric model with the
parameters from BGS.  
Contours levels are $n\times 0.1 \textrm{mag}^2, n\in I\!\!N$.}
\label{figcoberms}
\end{figure*}

\begin{figure*}
\resizebox{\hsize}{!}{\includegraphics{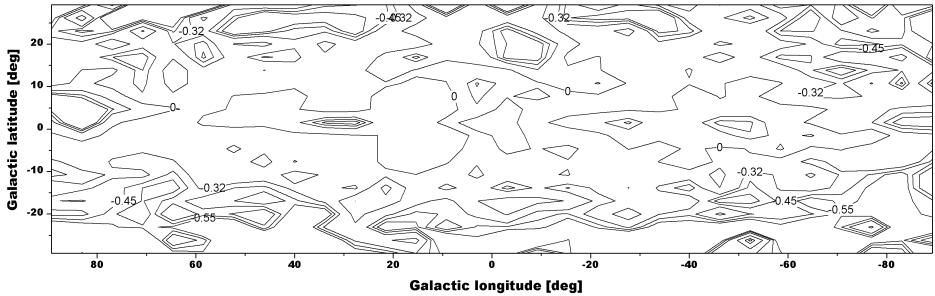}}
  \caption{Difference between {\it COBE/DIRBE} L-band data logarithmic
surface brightness (magnitudes) and the parametric model
from BGS in magnitudes. Negatives values for the contour
levels indicate that a model is too bright as compared to the data. The contour
levels are chosen such that their squares are equivalent to $\approx 0.1^m,
\approx 0.2^m {\rm and} \approx 0.3^m$, to allow a direct comparison with 
fig. \ref{figcoberms}.}
  \label{figsign}
\end{figure*}

As a first step we will investigate graphically 
whether an inhomogeneous variance pattern has to be assumed, 
which is indicated by inhomgeneous squares of residuals. 
Fig. \ref{figcobe} shows the {\it COBE/DIRBE} L-band data and figure
\ref{figcoberms} the residuals $r_{ij}^2 = \left(y_{ij} - 
\omega_{\hat \vartheta_{\rm BGS}}\right)^2$, with the 
model parameters $\hat \vartheta_{\rm BGS}$ taken from BGS. 
Provided this model holds (approximately) 
true, as an important conclusion from Fig. \ref{figcoberms} we find 
strong indication for inhomogeneous noise. Interestingly, towards the boundary 
of the observed part of the sky, the variability of the observations increases. 
Fig. \ref{figsign} shows the difference between the
model and the data including the algebraic sign of the difference. Note that
the error distribution is obviously inhomogeneous, both in the logarithmic
magnitude scale plotted in the figures and in a linear scale. 
Further note there is a systematic dependence of the sign of the deviations
on the position on the sky, whereas the model fits well in the central part
of the observed part of the sky.
This is indication that the MW disk shows deviations from an exponential
$\it z$-dependence.   

\begin{figure*}
\resizebox{\hsize}{!}{\includegraphics{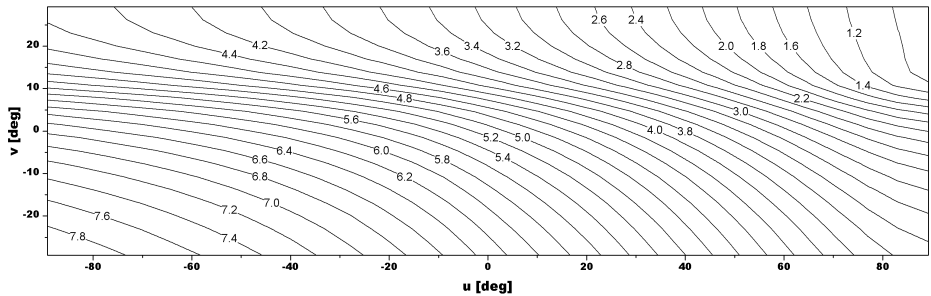}}
  \caption{ The smoothed {\it COBE/DIRBE} L-band data. This is the observed
data in the sky region $|l|\!\leq\! 60^o$ and  $-20^o\!\leq\! b\!\leq\! 10^o$,
after application of the smoothing operator {\bf T}. Contours indicated are for 
the natural logarithm of ${\bf T}\omega$. Note
that the $x$ and $y$ axis are from $|l|\!\leq\! 60^o$ and  
$-20^o\!\leq\! b\!\leq\! 10^o$, due to the definition of {\bf T}.}
\label{figcobesmooth}
\end{figure*}

We now determine the best-fit model parameter $\hat\theta^{\ast}$ by
minimisation of $\hat{D}^2$. We use the parameters found by BGS as starting values 
for our minimisation algorithm. 
Due to the increasing noise towards the boundary of the observed part of the sky, 
we restrict the region of the surface brightness data used in the fit 
to the region $|l|\!\leq\! 60^o$ and  $-20^o\!\leq\! b\!\leq\! 10^o$.
This is done to downweight those parts of the
sky where noninformative parts in the data are expected (see fig. \ref{figcoberms}).
Fig. \ref{figcobesmooth} shows this data after it has been smoothed with 
the smoothing operator $\bf T$. Note how much smoother the smoothed data appears
compared to the original {\it COBE/DIRBE} L-band data. 
Our computational  strategy consists of two steps. 
\begin{description}
\item[1. Fitting the disk:]
In the first step we fit the disk parameters with fixed bulge parameters.
\item[2. Fitting the bulge/bar:]
In the second step we fix
the disk related parameters found in the first step (except for the
normalisation parameter $d$) 
and fit the bulge/bar parameters and $d$.
\end{description}
Table \ref{table1} shows our result for the best-fit model parameters 
$\hat{\vartheta}^*$ and the model parameters of BGS. As suggested in sect.
\ref{bootstrap} we have obtained our best-fit model parameters by minimisation of $\hat{D}^2$. 
Note that BGS have not used exactly the same region of the sky in their fit 
as we use here. Therefore, one has to take into account that differences between 
these model parameters may be partially due to different  
regions of the sky (data) used in the fit. We reduce this problem by redetermining
the normalisations $b,d$ of the model by BGS (keeping fixed their other parameters)
with our algorithm, using the region of the sky selected above.
The value of our proposed statistical quantity $\alpha^{\ast}$ for the 
BGS model has been calculated for this modified version of their model parameters.

Applying the bootstrap algorithm presented in sect. \ref{bootstrap} to
our model we 
find that  $\alpha^{\ast}\!=\!0.86$, which indicates 
no significant  evidence against this model. 
For the parameters found in BGS 
a value $\alpha^{\ast}\!=\!0.80$  is obtained which yields a slightly worse 
fit. Note that, at a first glance, 
 this statement is  in  contradiction to the argument
given by BGS in the last paragraph of their p366. They pointed
out that a graphical inspection of residuals 
suggests that for the  model considered,
some local regions of the sky seem to
show systematic differences between their model and the observed data.
As a conclusion we find that the proposed 
method is not capable of concluding
that these local deviations between model and data  are due 
to systematic deviations. 
As pointed out by the referee this might be due to 
lack of power  of the proposed method,  
because an additional smoothing step was proposed. 
Indeed, this corresponds to some theoretical results 
concerning the asymptotic efficiency of the proposed method
(Munk \& Ruymgaart (1999)). 
 In fact, a more powerful method
could result from chosing a data-driven smoothing operator $\bf T$, similar 
to bandwidth selection in kernel regression.
  The main difficulty which arises is a different 
limit law compared to the case discussed
in the present, where $\bf T$ is fixed. 
However, this is beyond the scope
of this paper and will be an interesting topic for further research.

\begin{table*}
\begin{centering}
\begin{tabular}{lcccccccccc}
\hline
Parameters &
$z_0$ & $z_1$ & $\alpha$ & $r_d$ & $d$ &
$\eta$ & $\zeta$ & $a_m$ & $b$ & $\alpha^{*}$ \\
\hline
``Our'' &
$0.162$ & $0.042$ & $0.27$ & $2.56$ & $0.41$ &
$0.502$ & $0.59$ & $1.90$ & $306.1$ & $0.86$\\
``BGS'' &
$0.21$ & $0.042$ & $0.27$ & $2.5$ & $0.463$ &
$0.5$ & $0.6$ & $1.9$ & $234.4$ & $0.80$\\
\hline
\end{tabular}
\caption{\sl Parameter values for our model and the model
according to BGS and the statistical parameter $\alpha^{*}$
(\ref{eqsig}). Note that the parameters $\alpha$ and $z_1$ 
are not fitted}
\end{centering}
\label{table1}
\end{table*}

It can be seen from Table \ref{table1} that the main difference between the
two models is that our model has a lower disk scale height $z_0$.
The value $\alpha^{*}$ was found to be slightly  better for our new model compared to 
the BGS parameters. However, recall that 
we used only a part of the {\it COBE/DIRBE} L-band surface brightness
data in our fit.

\section{Results and conclusions}
We have 
argued that classical  measures of goodness of fit adopted from 
checking distributional assumptions 
can be misleading in the context of (inverse) regression. 
Particularly, an inhomogeneous  noise field can 
inflate the precision of common  $\chi^2$ quantities.
For this case, a new method was  proposed for noisy Fredholm eqs. of the first kind 
by Munk \& Ruymgaart (1999). As an example for the application of the
suggested algorithm, we use the problem of determining the luminosity density in the 
MW from surface brightness data. From this we have found that the parametric model in
Binney et al. (\cite{binney96}) can be improved slightly and gives
a satisfactory fit of the {\it COBE/DIRBE} L-band data in a range of 
$-20^o\!\leq\! b\!\leq\! 10^o$. 

\appendix
\section{Chosing the smoothing kernel $T$.} 
\label{remarks}
We mention that our procedure can also be performed with  any other  
smoothing kernel $T$. This will also yield  
in general different values of $\alpha^*$.
In principle, a valid option is any 
injective Operator $\bf T$. A good choice of $\bf T$, however, 
is driven by various aspects, such as efficiency or simplicity. 
An extensive simulation study  performed in Munk \& Ruymgaart (1999), 
reveals the kernel $T(u,v) = \min (u,v)$ as a reasonable choice which 
yields a procedure capable to detect a broad range of deviations from 
$U$. See, however, the discussion in sec. 5. 
A particularly simple choice  in noisy inverse models
\bea
y_i= {\bf K}f(t_i) + \varepsilon_i
\eea
can be achieved if 
{\bf T} is the adjoint of 
{\bf K} itself, provided
{\bf K} is a smoothing operator of the type
\bea
{\bf K }f(\cdot)=\int K(\cdot,v) f(v) d v.
\eea
However, in our application this is not easy to calculate and
will depend on constraints which force the particular model
$\rho_{\vartheta}$ to be identifiable.

\begin{acknowledgements}
The authors are indebted to the organizers F. Ruymgaart,
W. Stute and Y.Vardi  of the conference
'Statistics for Inverse Problems' held 
at the Mathematical Research Center at Oberwolfach, Germany, 1999.
The present paper   was essentially  initiated  by this meeting.
We would like to thank O. Gerhard and the referee, P. Saha, for 
many helpful comments.Nicolai Bissantz acknowledges support by 
the Swiss Science Foundation under grant number 20-56888.99.
\end{acknowledgements}


\begin{thebibliography}{99}
\bibitem[1999]{aerts}
Aerts, M.,  Claeskens, G., Hart, J., 1999, Journ. Americ. Statist. Assoc. 94, 869
\bibitem[1974]{akai}
Akaike, H., 1974, IEEE Trans. Autom. Control 19, 716
\bibitem[1997]{alcock}
Alcock, C., Allsman, R.A., Alves, D., et al., 1997, ApJ 486, 697
\bibitem[1993]{azzbow}
Azzalini, A.,  Bowman, A.W., 1993, Jour. Roy. Statist. Soc. Ser. B 55, 549
\bibitem[1984]{babu}
Babu, G.J., 1984, Shankya A 46, 85
\bibitem[1984]{barrow}
Barrow, J.D., Sonoda, D.H., Bhavsar, S.P., 1984, MNRAS 210, 19
\bibitem[1984]{vandenbergh}
van den Bergh, S., Morbey, C. L., 1984, ApJ 283, 598
\bibitem[1989]{berter}
Bertero, M., 1989, Adv. Electron. el. Phys. 75, 1
\bibitem[1994]{biboerner}
Bi, H., Boerner, G., 1994, A\&AS 108, 409
\bibitem[1996]{binney95}
Binney, J., Gerhard, O., 1996, MNRAS 279, 1005
\bibitem[1997]{binney96}
Binney, J.,  Gerhard, O., Spergel, D., 1997, MNRAS 288, 365 (BGS)
\bibitem[1997]{bissantz97}
Bissantz, N., Englmaier, P., Binney, J., et al., 1997, MNRAS 289, 651 
\bibitem[1997]{boazz}
Bowman, A.W.,
 Azzalini, A., 1997,
Applied smoothing 
techniques for data 
analysis: the kernel
 approach with S-Plus 
illustrations (Oxford Statistical 
Science Series. 18), Oxford: 
Oxford University Press.
\bibitem[1999]{cowei}
Cook, R.D., Weisberg, S., 1999, 
Applied Regression Including Computing and Graphics,
New York, NY: Wiley.  
\bibitem[1974]{cohi}
Cox, D.R., Hinkley, D.V., 1974, 
Problems and solutions in theoretical statistics,
A Halsted Press Book. London: Chapman and Hall. New York: John Wiley \& Sons.
\bibitem[1988]{cowa}
Cox, D., Koh, E., Wahba, G. Yandell, B.S., 1988,
Ann.  Statist. 18, 113
\bibitem[1997]{davhi}
Davison, A.C., Hinkley, D.V., 1997, 
Bootstrap methods and their application,
Cambridge Series on Statistical and Probabilistic Mathematics. Cambridge: Cambridge University Press. 
\bibitem[1998]{demu1}
Dette, H., Munk, A., 1998,  The Annals of Statistics 26, 778
\bibitem[1995]{dwek}
Dwek, E., Arendt, R.G., Hauser, M.G., et al., 1995,. Morphology, near-infrared luminosity, 
and mass of the Galactic bulge from COBE DIRBE observations, ApJ 445, 716
\bibitem[1979]{ef1}
Efron, B., 1979, Ann. Stat  7, 1
\bibitem[1993]{ef2}
Efron, B.,Tibshirani, R.J., 1993,
An Introduction to the Bootstrap, 
 Monographs on Statistics and Applied Probability. 57. New York, NY: Chapman \& Hall.
\bibitem[1992]{euha}
Eubank, R.L., Hart, J.D., 1992,  Ann. Statist. 20, 1412
\bibitem[1998]{freudenreich}
Freudenreich, H.T., 1998, ApJ  492, 495
\bibitem[1987]{gall}
Gallant, A.R., 1987, Nonlinear Statistical Models, 
Wiley Series in Prob. \& Mathem. Statist., Wiley: New York. 
\bibitem[1993]{haermam}
H\"ardle, W., Mammen, E., 1993, Ann. Statist.  21, 1926
\bibitem[1991]{haermar}
H\"ardle, W., Marron, J.S., 1991,
 Ann. Stat. 19, 778
\bibitem[1997]{hart}
Hart, J.D., 1997, Nonparametric smoothing and lack of fit tests,
Springer Series in Statistics. Springer.
\bibitem[1996]{hock}
Hocking, R.R., 1996,
Methods and Applications of Linear Models,
New York, NY: Wiley. 
\bibitem[1991]{kent}
Kent, S.M., Dame, T.M., Fazio, G., 1991, ApJ 378, 131
\bibitem[1994a]{lucy}
Lucy, L. B., 1994a, A\&A 289, 983
\bibitem[1994b]{lucy2}
Lucy, L. B., 1994b, Reviews in Modern Astronomy 7, 31
\bibitem[1997]{mair}
Mair, B.A., Ruymgaart, F.H., 1997,  SIAM J. Appl. Math. 56,  1424
\bibitem[1992]{matt}
Matthai, A.M., Provost, S.B., 1992, Quadratic forms in Random Variables,
Marcel Dekker, NY
\bibitem[1998]{mohdmok}
Mohdeb, Z., Mokkadem, A., 1998,
  C. R. Acad. Sci., Paris, Ser. I, Math  326, No.9, 1141
\bibitem[1992]{muel}
M\"uller, H.G., 1992, Scand. J. Statist. 19, 157 
\bibitem[1999]{mu1}
Munk, A., 1999, under revision, Scand. J. Stat.
\bibitem[1999]{muruy} Munk, A., Ruymgaart, F., 1999, submitted
\bibitem[1989]{nych}
Nychka, D.W.,  Cox, D.D., 1989,
 Ann. Stat.  17, 556
\bibitem[1994]{press}
Press, W.H., Teukolsky, S.A., Vetterling, W.T., et al., 1994,  Numerical recipes in C (2nd ed.),
Cambridge: Cambridge University Press.
\bibitem[1996]{rooi}
van Rooij, Ruymgaart, F.H., 1996, J. Statist. Plann.
Inference 53,  389
\bibitem[1995]{shaotu}
Shao, , J., Tu, D., 1995, The Jacknife and Bootstrap, 
Springer, NY.
\bibitem[1986]{shwe}
Shorack, G., Wellner, J., 1986, Empirical Process with Applications to 
Statistics, Wiley, NY.
\bibitem[1986]{simpson}
Simpson, G., Mayer-Hasselwander, H., 1986, A\&A 162, 340
\bibitem[1996]{spergel95}
Spergel, D.N., Malhotra, S., Blitz, L.: 1995,
Towards a Three-Dimensional Model of the Galaxy. In:  
ESO/MPA Workshop on Spiral Galaxies in
the Near-I, Minniti D., Rix H.-W. (eds,), Springer 1996
\bibitem[1998]{stut2}  Stute,W.,   Gonz\'ales-Manteiga, W., 
Presedo Quindimil, M.,
1998, Journ. Amer. Statist. Assoc. 93,
141
\bibitem[1995]{wajo}
Wand, M.P.; Jones, M.C., 1995, Kernel smoothing,
Monographs on Statistics and 
Applied Probability. 60. 
London: Chapman \& Hall.  
\bibitem[1994]{weiland94}
Weiland, J.L., et al., 1994,
ApJ 425, L81
\bibitem[1986]{wu}
Wu, C.F.J., 1986, 
 Ann. Stat.  14,
 1261
\end{thebibliography}
\end{document}